\documentclass[twocolumn,superscriptaddress,amsmath,amssymb]{revtex4-2}
\usepackage{color}
\usepackage{graphicx}
\usepackage{subfigure}
\usepackage{booktabs}
\usepackage{dcolumn}
\usepackage{bm}
\usepackage{amsmath}
\usepackage{verbatim}
\usepackage{lineno}

\usepackage{placeins} 

\newcommand{\beq}{\begin{eqnarray}}
\newcommand{\eeq}{\end{eqnarray}}

\begin{document}

\title{Relaxation of quenched structural glasses: descent in a stiffening caging potential over inflection‑point `speed bumps'}

\author{Ting Qu}
\affiliation{Institute of Theoretical Physics, Chinese Academy of Sciences, Beijing 100190, China}
\affiliation{School of Physical Sciences, University of Chinese Academy of Sciences, Beijing 100049, China}

\author{Deng Pan}
\affiliation{Institute of Theoretical Physics, Chinese Academy of Sciences, Beijing 100190, China}


\author{Yuliang Jin}
\email{yuliangjin@mail.itp.ac.cn}
\affiliation{Institute of Theoretical Physics, Chinese Academy of Sciences, Beijing 100190, China}
\affiliation{School of Physical Sciences, University of Chinese Academy of Sciences, Beijing 100049, China}
\affiliation{Wenzhou Institute, University of Chinese Academy of Sciences, Wenzhou, Zhejiang 325000, China}

\date{\today}

\begin{abstract}
The slow energy relaxation  in quenched glasses is a ubiquitous yet poorly understood phenomenon. Despite extensive study, the microscopic origin of the observed power-law decay remains debated, with proposed mechanisms ranging from saddle-point slowdown and marginal stability to coarsening of localized excitations and phonon dynamics. Here, by simulating gradient descent in archetypal structural glass formers,  we show that none of these scenarios can account for our data. Instead, the power-law behavior emerges from a remarkably simple caging effect: each particle experiences an effective stiffening potential that arises from many-body confinement and
diverges at a characteristic cage size. 
This mechanism analytically yields the observed power-law decay and is quantitatively reproduced by a minimal single-particle cage model with fixed neighbours, demonstrating that collective relaxation modes are not essential.
The dynamics is punctuated by fluctuations as the system rolls through inflection points on the energy landscape, which act as `speed bumps' but do not affect the overall power-law behaviour. In contrast to mean-field spin glass theory, we find no characteristic temperature that separates distinct dynamical regimes; state following within a given glass basin occurs universally for all initial temperatures whenever the system is sufficiently close to the inherent structure. Our results establish a complete physical picture of gradient descent dynamics in typical structural glasses.
\end{abstract}

\maketitle
{\bf Introduction.}
Glasses are made by suddenly quenching a high-temperature melt to low temperatures~\cite{angell1995formation}. The relaxation of glass is a paradigm of far-from-equilibrium dynamics in disordered systems, during which the system minimizes its energy under significant influence from 
the inherent disorder. Other important examples of such non-equilibrium relaxation dynamics include the training of artificial neural networks in machine learning~\cite{baity2018comparing, huang2025liquid} and the folding pathways of proteins~\cite{frauenfelder1991energy}.

The relaxation of a quenched glass  is mathematically described by a gradient descent (GD) equation~\cite{kurchan1996phase, folena2020rethinking, chacko2019slow, ikeda2020universal, nishikawa2022relaxation, shimada2024instantaneous, manacorda2022gradient}:
\begin{equation}
\frac{d {\bf r}_i}{dt} = - \frac{\partial V(\{ {\bf r}_i\})}{\partial {\bf r}_i},
\label{eq:GD}
\end{equation}
where ${\bf r}_i$ are the coordinates of particle $i$ (in spin glasses, ${\bf r}_i$ is replaced by the spin variable $\sigma_i$), and $V(\{ {\bf r}_i\})$ is the total potential energy. Equation~(\ref{eq:GD}) is a Langevin equation with zero noise, assuming the final temperature is sufficiently low that thermal fluctuations are negligible (the damping coefficient is set to unity). Typically, the initial condition is an equilibrium configuration at temperature $T$, sampled according to the Boltzmann distribution, and the final state is a stable (or marginally stable) configuration at zero temperature. 

If the system's energy 
is simply a 
one-dimensional (1D) harmonic potential, $V = \frac{1}{2}\lambda r^2$, then the solution of Eq.~(\ref{eq:GD}) yields exponential decay: $r(t) \sim e^{-\lambda t}$ and $V(t) \sim e^{-2\lambda t}$. However, in disordered systems, $V({ \mathbf{r}_i})$ generally represents a complex energy landscape composed of numerous metastable minima and the pathways leading to them. Consequently, the GD dynamics becomes highly complex.

In many glassy systems, the relaxation of the sample-averaged energy, $E = \langle V \rangle$, exhibits power-law relaxation:
\begin{equation}
E(t) - E_\infty \sim t^{-\alpha},
\label{eq:E_relax}
\end{equation}
where $E_\infty = E(t \to \infty)$.
Such systems include mean-field spherical pure~\cite{kurchan1996phase} and mixed~\cite{folena2020rethinking} $p$-spin glasses, mean-field spheres with Mari–Kurchan (MK) interactions~\cite{nishikawa2022relaxation}, harmonic and soft spheres~\cite{chacko2019slow, ikeda2020universal, saitoh2020stress, nishikawa2022relaxation, interiano2024critical}, the Kob–Andersen Lennard-Jones (KALJ)  model~\cite{nishikawa2022relaxation, shimada2024instantaneous}, and the random Lorentz gas~\cite{manacorda2022gradient}.

Equation~(\ref{eq:E_relax}) is equivalent to power-law relaxation of the mean velocity:
\begin{equation}
v(t) \sim t^{-\beta},
\label{eq:v_relax}
\end{equation}
where $v(t)
=\left\langle\sqrt{
\frac{1}{N}
\sum_i
\left|\dot{\mathbf r}_{i}(t)\right|^2}
\right\rangle$ and $\alpha = 2\beta - 1$. Equation~(\ref{eq:v_relax}) is numerically more convenient because it does not require estimating the asymptotic long-time value.
For structural glasses, the reported value of $\beta$  ranges from $0.75$ to $1.25$, across various models and dimensions~\cite{folena2020rethinking, nishikawa2022relaxation, chacko2019slow, ikeda2020universal, manacorda2022gradient, shimada2024instantaneous, interiano2024critical} (i.e., $\alpha \in [0.5, 1.5]$).

\begin{figure}[!htbp]
  \centering
  \includegraphics[width=\linewidth] {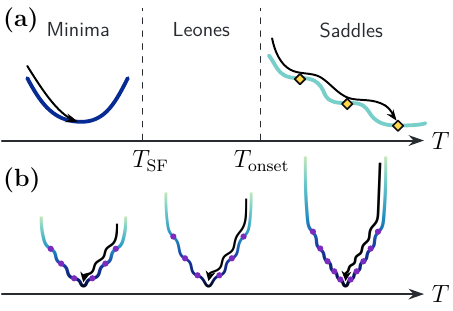}
\caption{{\bf  
Two pictures for GD dynamics in glasses.}
(a) Mean-field mixed $p$-spin picture~\cite{folena2020rethinking, folena2023introduction}. For $T > T_{\rm onset}$, the dynamics is dominated by saddle points (yellow diamonds) before ending in a marginal state. For $T < T_{\rm SF}$, the dynamics follows a state within a basin. The regime $T_{\rm SF} < T < T_{\rm onset}$ is not fully understood theoretically.
(b) Picture for structural glasses. The system descends in a stiffening caging potential, passing through inflection points (purple circles). The starting point lowers with decreasing $T$, but no characteristic temperature separates qualitatively distinct dynamical regimes.
}
\label{fig:schematic}
\end{figure}

\begin{figure*}[!htbp]
  \centering
  \includegraphics[width=\linewidth] {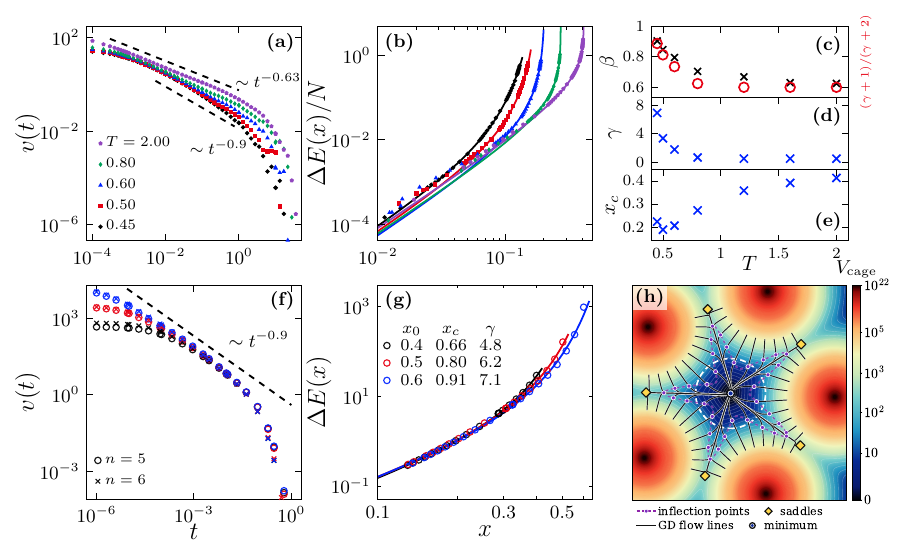}
\caption{ {\bf Gradient descent dynamics and the effective caging potential.}
(a-e) Results for the 3D KALJ model. (a) Average velocity $v(t)$ for different temperatures $T$, with power-law fits yielding $\beta$. 
(b) Average energy $\Delta E(x)$ per particle for different $T$, fitted to Eq.~(\ref{eq:Vcage}) (solid lines), with the  fitting parameters $\gamma$ and $x_c$ given in (d, e). 
(c) Comparison between $\beta$ from the fits in (a) (black crosses) and $(\gamma+1)/(\gamma+2)$ (red circles), where $\gamma$ is obtained from the fits in (b).
(f-h) Results for the 2D MCM. (f) $v(t)$ and (g) $\Delta E(x)$ for a few different parameter settings.
The solid lines in (g) are fits to Eq.~(\ref{eq:Vcage}). The dashed line in (f) indicates $\beta = 0.9$ obtained from the relation $\beta = (\gamma+1)/(\gamma+2) \approx 0.9$.
(h) Energy landscape of a cage with $n=5$ neighbors. 
}
\label{fig:cage_potential_KALJ}
\end{figure*}

\begin{figure}[!htbp]
  \centering
  \includegraphics[width=\linewidth] {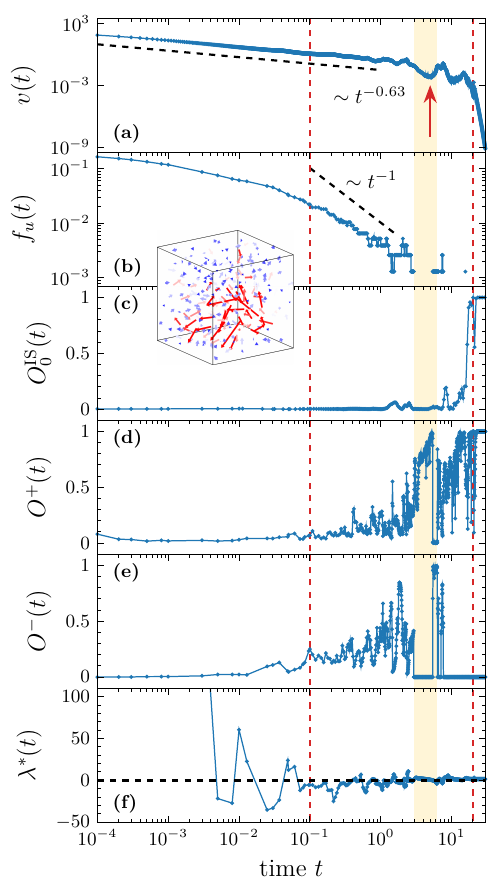}
\caption{ {\bf Gradient descent dynamics in a typical KALJ sample.}
All quantities are plotted as functions of $t$.
(a) Mean velocity per particle $v(t)$. 
(b) Fraction $f_u(t)$ of unstable instantaneous normal modes.
(c) Overlap $O_0^{\rm IS}(t)$ between the normalized velocity field $\hat{{\bf v}}(t)$ and the softest vibrational mode of the inherent structure. 
(d) Largest overlap $O^+$ between $\hat{{\bf v}}(t)$ and stable instantaneous normal modes. 
(e) Largest overlap $O^-$ between $\hat{{\bf v}}(t)$ and unstable instantaneous normal modes.
(f) Eigenvalue $\lambda^*$ of the instantaneous normal mode $\mathbf{x}_{n^*}$ that has the largest overlap with $\hat{\mathbf{v}}(t)$. 
The vertical dashed lines mark $\tau_{\rm IP} \approx 0.1$ and $\tau_{\rm harm} \approx 20$. The yellow band and the red arrow indicate an inflection-point event, analyzed in detail in Fig.~\ref{fig:inflection-point}.
The inset shows the softest mode after $\tau_{\rm harm}$; large (small) displacements are colored in red (blue).}
\label{fig:relaxation_KALJ}
\end{figure}

Understanding the physical origin of this power-law relaxation has been central to many recent studies. The following four mechanisms have been proposed:

(i) {\it Saddle-point slowdown and marginal stability of glass states}.
The connection between power-law relaxation and the marginal property of glass states arises from the dynamical solutions of mean-field spherical  $p$-spin glasses~\cite{kurchan1996phase, folena2020rethinking}. Recently, Ref.~\cite{folena2020rethinking} solved the dynamical mean-field equations for spherical mixed $p$-spin glass models and presented the following picture (see Fig.~\ref{fig:schematic}a).  (1) If the initial temperature $T$ is above a characteristic onset temperature $T_{\rm onset}$, the final energy $E_\infty = E_{\rm th}$ is independent of $T$, with $\alpha_{\rm MFSG} = 2/3$ (or $\beta_{\rm MFSG} \approx 0.83$). The dynamics slows down as it passes near saddle points before reaching a marginally stable configuration.
(2) When $T_{\rm SF} < T < T_{\rm onset}$, where $T_{\rm SF}$ is a characteristic temperature for state-following dynamics, the power-law form of Eq.~(\ref{eq:E_relax}) still holds with the same exponent $\alpha_{\rm MFSG} = 2/3$, but $E_\infty$ decreases with $T$.
This regime is called {\it hic sunt leones} in Ref.~\cite{folena2020rethinking} and remains poorly understood even for mean-field spin glasses.
(3) For $T < T_{\rm SF}$, the energy $E(t)$ decays exponentially because the initial configuration lies very close to a minimum, and the system remains in the same glass state throughout relaxation — a regime known as {\it state-following}~\cite{franz1995recipes, rainone2015following, zdeborova2010generalization}.

The mean-field spin glass mechanism of power-law relaxation can be understood through a simple argument based on the properties of the Hessian matrix $H$, given in~\cite{folena2023introduction}. 
For spherical $p$-spin glasses, the Hessian is a random matrix whose  density of states (DOS) $\rho(\lambda)$ follows a Wigner semicircle distribution that includes small negative eigenvalues. Near the minimum eigenvalue $\lambda_{\rm min}(t) \sim E_\infty - E(t) = -\Delta E(t) < 0$,
the DOS follows $\rho(\lambda) \sim \sqrt{\lambda - \lambda_{\rm min}}$. The presence of negative eigenvalues indicates that the fixed points are saddles. Assuming the dynamics randomly explores all eigenvector directions, the time required to escape the vicinity of a saddle should be inversely proportional to the fraction $f_u$ of negative eigenvalues. Since $f_u = \rho(\lambda < 0) \sim \int_{\lambda_{\rm min}}^{0} d\lambda \, \sqrt{\lambda - \lambda_{\rm min}} \sim \Delta E^{3/2}$, the escape time scales as $t \sim 1/f_u \sim \Delta E^{-3/2}$, which yields $\alpha_{\rm MFSG} = 2/3$. Notably, in the limit $t \to \infty$, $\rho(\lambda) \sim \sqrt{\lambda}$ becomes pseudogapped, a signature of marginal states. Saddle-point slowdown and marginality are thus the origins of power-law relaxation in this picture.

It remains unclear whether the above mechanism applies to structural glasses in finite dimensions~\cite{nishikawa2022relaxation, shimada2024instantaneous}. Even for the mean-field MK model, the measured exponent $\alpha_{\rm MK} = 0.5$ ($\beta_{\rm MK} = 0.75$) differs from $\alpha_{\rm MFSG} = 2/3$, suggesting a lack of universality~\cite{nishikawa2022relaxation}. A recent study has attempted to connect the power-law relaxation in the KALJ model to the unstable instantaneous normal modes during the GD process~\cite{shimada2024instantaneous}.

(ii) {\it Coarsening of localized excitations.}
From a real-space perspective, power-law relaxation may arise from the slow coarsening of ``hot spots'', which are localized regions of particles exhibiting large non-affine displacements. For over-jammed harmonic spheres, Ref.~\cite{chacko2019slow} finds a power-law growth of the correlation length (coarsening length) between hot spots, $l^* \sim t^{\gamma_l}$, where $\gamma_l = 0.37$ in 2D and $\gamma_l = 0.25$ in 3D.

(iii) {\it Phonon excitations.}
At sufficiently low $T$, one expects only phonons to be excited during the GD process~\cite{nishikawa2022relaxation}. According to the Debye law, the phononic DOS in $D$ dimensions is $\rho_{\rm phonon}(\omega) \sim \omega^{D-1}$, or equivalently 
$\rho_{\rm phonon}(\lambda) \sim \lambda^{D/2-1}$, where $\lambda = \omega^2$. 
Since each phonon mode is harmonic, the corresponding energy decays as $e^{-2\lambda t}$. The total energy decay is then given by $\Delta E \sim \int d\lambda \, \rho_{\rm phonon}(\lambda) \, e^{-2\lambda t} \sim t^{-D/2}$, yielding $\alpha_{\rm phonon} = D/2$ and $\beta_{\rm phonon} = D/4 + 1/2$. The existence of quasi-localized low-energy excitations at low $T$, with $\rho_{\rm QLE}(\omega) \sim \omega^4$, does not modify the exponent $\alpha_{\rm phonon}$, since $\omega^4$ is a higher-order contribution to $\rho(\omega)$~\cite{nishikawa2022relaxation}.

(iv) {\it Critical-like slowdown near the jamming transition.} Near the jamming transition of harmonic spheres, the energy behaves as $E(t) \sim t^{-\alpha_{\rm J}} e^{-t/t^*}$, with $\alpha_{\rm J} \approx 1$ and $t^*$ diverging at the jamming transition~\cite{ikeda2020universal, interiano2024critical}. This behavior is reminiscent of critical  slowdown near a critical point.

This study provides a concrete understanding of the GD relaxation process in structural glasses. First, we find that all four existing scenarios (i-iv) are inapplicable. Instead, the observed power-law behavior can be explained by a simple mechanism: an effective caging potential with divergent stiffening effects. A minimal single-cage model suffices to reproduce the observed power law. Second, we show that the fluctuations around the power-law decay are caused by the system passing through inflection points on the energy landscape, which act as speed bumps during the descent. Third, by analyzing the trajectory's response to perturbations, we find that a state-following regime exists at any temperature $T$, as long as the system is sufficiently close to the inherent structure (minimum). The boundary of this state-following regime depends on the perturbation magnitude and the distance to the inherent structure, but it is universal across all studied $T$.

Our key findings are summarized by the schematic plot in Fig.~\ref{fig:schematic}b. The GD dynamics in structural glasses is essentially described as descent within a stiffening caging potential, passing through inflection-point speed bumps that cause velocity fluctuations.
In sharp contrast to the mean-field spin glass picture (Fig.~\ref{fig:schematic}a), the GD dynamics in structural glasses is not influenced by saddle points or marginal states. The dynamics are qualitatively similar at all temperatures, and no characteristic temperature can be defined to separate a low-$T$ state-following regime.



\begin{figure}[!htbp]
  \centering
  \includegraphics[width=\linewidth] {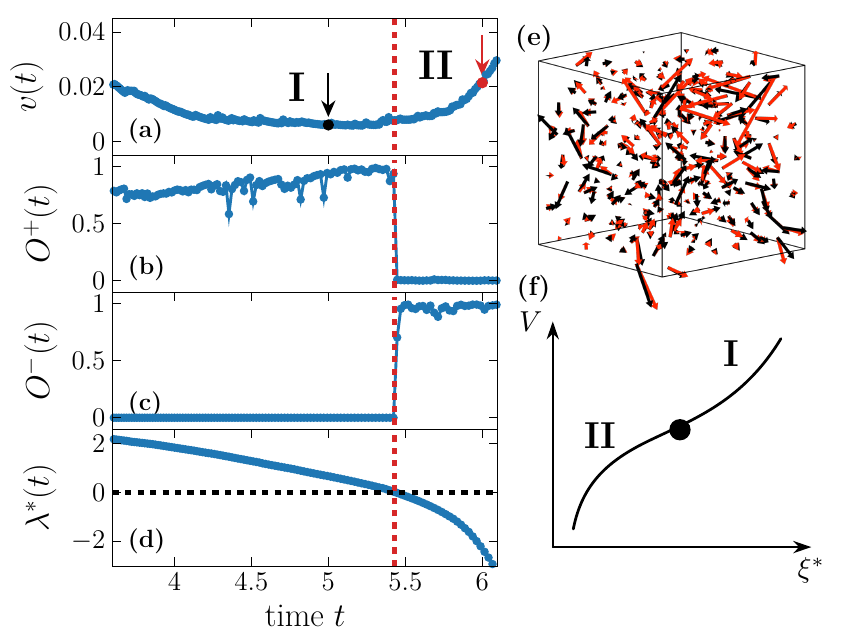}
\caption{{\bf Rolling through an inflection point on the energy landscape.} The event highlighted by the arrow in Fig.~\ref{fig:relaxation_KALJ}a is analyzed in detail here. (a) Mean velocity $v(t)$. (b) Largest overlap $O^+$ between $\hat{\mathbf{v}}(t)$ and the stable instantaneous normal mode $\mathbf{x}_{n^*_+}$. (c) Largest overlap $O^-$ between $\hat{\mathbf{v}}(t)$ and the unstable instantaneous normal mode $\mathbf{x}_{n^*_-}$. (d) Eigenvalue $\lambda^*$ of the instantaneous normal mode $\mathbf{x}_{n^*}$ that has the largest overlap with $\hat{\mathbf{v}}(t)$. (e) Visualization of the modes with the largest overlap: $\mathbf{x}_{n^*_+}$ (black) and $\mathbf{x}_{n^*_-}$ (red), at the marked times (arrows) in (a).  (f) Schematic potential energy $V$ as a function of the mode coordinate $\xi^*$ along the direction of $\mathbf{x}_{n^*}$. At the inflection point (the black point), the first derivative of the energy (the velocity or force) is minimum, while the second derivative is zero (the relevant eigenvalue $\lambda^*$ of the Hessian).
}
\label{fig:inflection-point}
\end{figure}

\begin{figure}[!htbp]
  \centering
  \includegraphics[width=\linewidth] {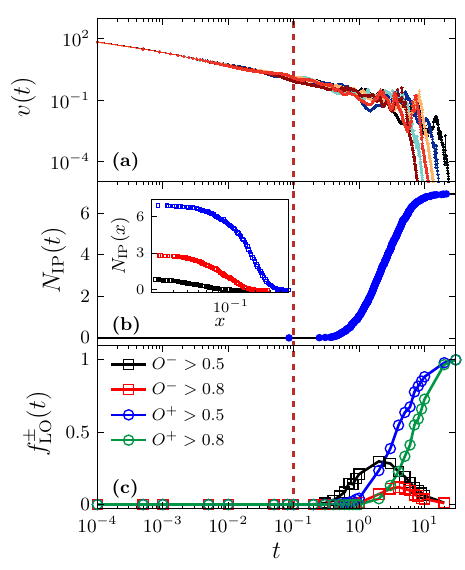}
\caption{{\bf Statistics of inflection-point events.} (a) Velocity $v(t)$ for ${10}$ independent samples for $T=2$. 
(b) Cumulative number $N_{\rm IP}(t)$ of inflection-point events, averaged over ${100}$ samples. 
(c) Fraction $f_{\rm LO}^{\pm}(t)$ of samples that satisfy $O^{\pm}(t) > O_{\rm threshold}$, where $O^{\pm}(t)$ is the largest overlap between the normalized velocity field $\hat{\mathbf{v}}(t)$ and stable/unstable instantaneous modes
(in total, 600 samples are analyzed).
The vertical line marks $\tau_{\rm IP} \approx {0.1}$.
Inset: $N_{\rm IP}(x)$ for  $T={2.0,0.6,0.45}$ (from top to bottom), where $x$ is the distance from the inherent structure; the characteristic distance $x_{\rm IP}$ decreases with decreasing $T$.}
\label{fig:mult_samples}
\end{figure}

{\bf Examination of existing mechanisms for the power-law relaxation.}
We simulate GD dynamics, Eq.~(\ref{eq:GD}), for two representative glass models: the 3D KALJ and amorphous silica models (see Methods for details). The overall behavior of $E(t)$ appears similar in the two models. 
The results presented in the main text are for the KALJ model; results for amorphous silica can be found in the Supplementary Information.

The mechanisms (iii) and (iv) can be easily ruled out as the origin of power-law relaxation in our systems. The simulated models do not exhibit a jamming transition or any other athermal critical points, so they should not display the critical‑like slowdown described in (iv). Moreover, according to Ref.~\cite{nishikawa2022relaxation},  the phonon mechanism (iii) applies only for \(T \ll T_{\rm def}\), where \(T_{\rm def} \approx 0.6\) (for KALJ) is a characteristic temperature below which particle rearrangements during minimization are significantly reduced (see Supplementary Section S1 for the determination of $T_{\rm def}$). However, we observe power‑law relaxation for all temperatures in the range \(0.4 \leq T \leq 2.0\) (see Fig.~\ref{fig:cage_potential_KALJ}a), which covers the onset temperature \(T_{\rm onset} \approx 0.7\) (see Supplementary Section S2 for the estimation of $T_{\rm onset}$). 
The power‑law behavior over such a wide range of \(T\) simply cannot be explained by the phonon mechanism.

Next, we examine the coarsening mechanism (ii), and find that the coarsening length $l^*$ grows logarithmically, $l^* \sim \ln t$, instead of following a power-law (see Supplementary Section S3). In fact, $l^*$ increases by only about a factor of three when $t$ spans over three decades. This logarithmic growth contrasts with the power-law growth $l^* \sim t^{\gamma_l}$ reported in Ref.~\cite{chacko2019slow}. A possible reason for this discrepancy is that granular models near the jamming transition belong to a different universality class possessing hierarchical  energy scales
~\cite{wu2026hierarchical}, whereas  the KALJ model lacks such properties. Note that logarithmic coarsening has already been reported for  a binary LJ mixture~\cite{oku2020phase}.

Finally, we examine mechanism (i). Figure~\ref{fig:relaxation_KALJ}b shows the fraction $f_u(t)$ of negative eigenvalues of the Hessian matrix of the instantaneous configuration at time $t$, obtained for a typical sample with $T=2$. The data do not follow $f_u(t) \sim t^{-1}$, as suggested by mean-field spin glass theory. In the interval $10^{-4} < t < 10^{0}$, we observe a robust power-law relaxation $v(t) \sim t^{-\beta}$ with a constant exponent {$\beta \approx 0.63$} (Fig.~\ref{fig:relaxation_KALJ}a). However, during this same time interval, any attempt to fit $f_u(t)$ to a power law yields an exponent that clearly varies with $t$. Consequently, there is no direct correspondence between the power-law decay of $v(t)$ and the behavior of $f_u(t)$, implying that the observed power law cannot arise purely from relaxation along unstable modes. Further evidence will be presented later to support the absence of saddle-point influence during the GD dynamics.

{\bf An effective stiffening caging potential for power-law relaxation.}
Since none of the existing scenarios are applicable, a new mechanism is needed to explain the observed power law. We find that a surprisingly simple mechanism — the glass caging effect — is sufficient to explain this behavior. This effect is characterized by an effective, stiffening caging potential, as revealed by the analysis of simulation results. We then construct a minimal cage model to reproduce the power-law relaxation.

Asymptotically, the system approaches a local energy minimum, whose configuration is called an inherent structure. We define the average distance $x(t)$ to the inherent structure as
\(
x^2(t) = \left\langle \Delta r^2(t) \right\rangle = \left\langle \frac{1}{N} \sum_i |\mathbf{r}_i(t) - \mathbf{r}_i^{\rm IS}|^2 \right\rangle,
\)
where $\langle \cdots \rangle$ denotes an average over samples, and $\{ \mathbf{r}_i^{\rm IS} \} = \{ \mathbf{r}_i(t \to \infty) \}$ is the inherent structure. Figure~\ref{fig:cage_potential_KALJ}b shows a parametric plot of $\Delta E(t)$ versus $x(t)$. The curves display harmonic behavior, $\Delta E \sim x^2$, for small $x$. At larger $x$, a rapid growth of $E(x)$ is observed, resembling an effective stiffening effect seen in certain soft materials, such as stretched polymers~\cite{doi1996introduction}, granular materials~\cite{pan2023shear}, and soft composites~\cite{zhao2026programming}.

The overall behavior of $E(x)$ can be well described by a phenomenological potential (Fig.~\ref{fig:cage_potential_KALJ}b):
\begin{equation}
V_{\rm cage}(x) = A \left[ \left( x_c - x \right)^{-\gamma} - \gamma \, x_c^{-\gamma-1} x - x_c^{-\gamma} \right],
\label{eq:Vcage}
\end{equation}
where $A$, $x_c$, and $\gamma$ are fitting parameters. The last two terms remove the constant and the first derivative of the potential at $x = 0$, ensuring that the system is stable. The divergence term $\left( x_c - x \right)^{-\gamma}$ implies that particles are confined within cages of a characteristic size $x_c$. These cages are formed by neighboring particles, and the divergence arises from the strong repulsion when the central particle approaches its neighbors.

With the effective potential Eq.~(\ref{eq:Vcage}), two asymptotic limits of the solution of Eq.~(\ref{eq:GD}) can be obtained analytically.

(i) In the limit $t \to \infty$ and $x \to 0$, $V_{\rm cage} \approx \frac{1}{2}\lambda x^2$, with $\lambda = A \gamma (\gamma+1) / x_c^{\gamma+2}$, and $E(t)$ decays exponentially, $E(t) \sim e^{-2\lambda t}$. 

(ii) In the limit $x \to x_c^-$, Eq.~(\ref{eq:Vcage}) approximates to $V_{\rm cage} \approx A (x_c - x)^{-\gamma}$ (neglecting the constant term). The dynamics starts from $x(t=0) = x_c - \delta x$, where $\delta x$ is a small positive parameter. Integrating Eq.~(\ref{eq:GD}) yields
\(
[x_c - x(t)]^{\gamma+2} - \delta x^{\gamma+2} = \gamma(\gamma+2) A t.
\)
For very small $t$ such that $\gamma(\gamma+2) A t \ll \delta x^{\gamma+2}$, the solution reduces to a constant independent of $t$, $x(t) \approx x_c - \delta x$,  which explains the plateau in $v(t)$ at very short times. As $t$ increases, the solution crosses over to the power-law decay $v(t) \sim t^{-\beta}$ with $\beta = (\gamma+1)/(\gamma+2)$. This relationship is confirmed by our simulation data, using the two exponents $\gamma$ and $\beta$ obtained from fits at different $T$ (see Fig.~\ref{fig:cage_potential_KALJ}c,d).

The parameter $x_c$ is plotted in Fig.~\ref{fig:cage_potential_KALJ}e. Its value falls in the range ${0.18} \leq x_c \leq {0.41}$ over the temperature interval ${0.45} \leq T \leq {2.0}$, and remains below the  nearest-neighbor distance $R_1 \sim 1$, 
which is given by the position of the first peak in the pair correlation function $g(r)$. This confirms that $x_c$ can be interpreted as an effective cage size, as it is within the first coordination shell.

The original GD problem is rather complicated: it describes the time evolution of a many-body system with $3N$-dimensional coordinates $\{\mathbf{r}_i(t)\}$ moving in a complex energy landscape $V(\{\mathbf{r}_i(t)\})$. Equation~(\ref{eq:Vcage}) greatly simplifies the description by reducing the $3N$-dimensional dynamics to a single one-dimensional coordinate $x$. The key insight is that the power-law relaxation observed in GD dynamics arises from the divergent stiffening term $(x_c - x)^{-\gamma}$ in this  effective cage potential.

{\bf A minimum cage model.} The caging potential description suggests that the overall dynamics is dominated by particles descending within the cages formed by their neighbors. Collective motions of the neighbors and other particles, although present, are not key factors. To support this idea, we propose a {\it minimum cage model} (MCM) that describes the GD dynamics of a single particle within a fixed cage. Despite its simplicity, the model reproduces the behavior of $E(t)$ well.

Without loss of generality, we consider the MCM in 2D, where $n$ neighboring particles are placed on a circle of radius $R_1 = 1.2$ to form a cage. The neighboring particles are fixed, while the mobile central particle interacts with them through a truncated LJ potential. The mobile particle is released from a random initial position on a circle of radius $x_0$ and then follows the GD dynamics defined by Eq.~(\ref{eq:GD}) (see Methods Appendix C for further details). 

The dynamics is described by a trajectory  in a 2D energy landscape. Figure~\ref{fig:cage_potential_KALJ}h shows an example of such a landscape, indicating the minimum, saddle points and typical GD flow lines. The resulting average energies $\Delta E(x)$ and $\Delta E(t)$ closely resemble those measured in the original problem (see Fig.~\ref{fig:cage_potential_KALJ}). Despite its simplicity, the exponent $\beta \approx 0.9$ obtained from this MCM is in line with the reported value $0.92 \leq \beta \leq 1$ for the 2D KALJ model~\cite{nishikawa2022relaxation}. Different choices of the parameters, such as $x_0$ and $n$, do not alter the essential power-law relaxation behavior.

{\bf Long-time dynamics: harmonic limit.} 
With the overall power-law behavior accounted for, we now examine two additional features of the GD dynamics that are most visible in single-sample data (Fig.~\ref{fig:relaxation_KALJ}): the exponential decay following the power-law regime and the late-stage fluctuations in the power-law regime. We address the former first.

Near the inherent structure, the potential in Eq.~(\ref{eq:GD}) can be approximated harmonically as $V(\Delta \mathbf{r}) \approx V(0) + \frac{1}{2} \Delta \mathbf{r}^{\mathsf{T}} H \Delta \mathbf{r}$, where $\Delta \mathbf{r} = \{\Delta \mathbf{r}_i\}$ denotes the particle displacements from the inherent structure. Equation~(\ref{eq:GD}) then reduces to a linearized form,
$\frac{d \Delta \mathbf{r}}{dt} = - H \Delta \mathbf{r}$.
Expanding $\Delta \mathbf{r}(t)$ in terms of the eigenvectors $\mathbf{x}_n$ of the Hessian $H$ as $\Delta \mathbf{r}(t) = \sum_n c_n(t) \mathbf{x}_n$, we obtain $c_n(t) = c_n(0) e^{-\lambda_n t}$ with $\lambda_n$ the corresponding eigenvalue, analogous to a 1D harmonic oscillator. Consequently, the velocity field $\mathbf{v}(t) = d\Delta \mathbf{r}(t)/dt =
\sum_n \dot{c}_n(t) \mathbf{x}_n$, 
also exhibits exponential decay.


The velocity field $\mathbf{v}(t)$ can be measured directly in our simulations (the  normalized   velocity field is $\hat{\mathbf{v}}(t) = \frac{\mathbf{v}(t)}{\|\mathbf{v}(t)\|} $).  To avoid confusion, below we denote the Hessian of an instantaneous configuration at time $t$ by $H(t)$, and the Hessian of the inherent structure by $H^{\rm IS} = H(t=\infty)$. The coefficients $q_n^{\rm IS}(t)$ in the expansion, 
{$\hat{\mathbf{v}}(t) = \sum_n q_n^{\rm IS}(t) \mathbf{x}_n^{\rm IS}$},
are obtained from $q_n^{\rm IS}(t) = \hat{\mathbf{v}}(t) \cdot \mathbf{x}_n^{\rm IS}$. Because both $\hat{\mathbf{v}}(t)$ and $\mathbf{x}_n^{\rm IS}$ are unit vectors, the coefficients satisfy the normalization condition $\sum_n |q_n^{\rm IS}(t)|^2 = 1$.

Figure~\ref{fig:relaxation_KALJ}c shows the overlap 
{$O_0^{\rm IS}(t) = |q_0^{\rm IS}(t)|^2$} between $\hat{\mathbf{v}}(t)$ and the softest mode $\mathbf{x}_0^{\rm IS}$, i.e., the eigenvector with the smallest eigenvalue. For the sample shown, the overlap reaches unity at $t = \tau_{\rm harm} \approx 20$, after which $v(t)$ decays exponentially (see Fig.~\ref{fig:relaxation_KALJ}a). This exponential tail confirms our earlier interpretation: the basin of the glass is stable near its bottom, rather than marginally stable, which would be accompanied by power-law asymptotic decay. Moreover, the relaxation of $v(t)$ is strongly coupled to the softest mode, which is spatially disordered (Fig.~\ref{fig:relaxation_KALJ}, inset). 
Coupling to modes with larger eigenvalues $\lambda_n$ decays much faster because of the exponential form of the coefficient $c_n \sim e^{-\lambda_n t}$.
The observed behavior differs from the phonon mechanism discussed in the Introduction, where $v(t)$ is expected to couple to all phonon modes, not only to the lowest-energy one. The sample-to-sample fluctuations of $\tau_{\rm harm}$ can be observed in Fig.~\ref{fig:mult_samples}a.

{\bf Inflection points as ``speed bumps''.}
Prior to $\tau_{\rm harm}$, the coupling of $\hat{\mathbf{v}}(t)$  to the softest mode of the inherent structure is small (small $O_0^{\rm IS}(t)$). It can be shown that for $\tau_{\rm IP} < t < \tau_{\rm harm}$, $\hat{\mathbf{v}}(t)$ instead couples to instantaneous normal modes and exhibits large fluctuations, whereas for $t < \tau_{\rm IP}$ this coupling is negligible. To quantify this behavior, we introduce two additional overlap parameters, $O^\pm(t) = |q_{n^*_\pm}(t)|^2$, where $q_{n^*_\pm}(t) = \hat{\mathbf{v}}(t) \cdot \mathbf{x}_{n^*_\pm}(t)$. Here
$\mathbf{x}_{n^*_+}(t)$ and $\mathbf{x}_{n^*_-}(t)$ denote the instantaneous normal modes that have the largest overlap with $\hat{\mathbf{v}}(t)$ among those with positive and negative eigenvalues, respectively.

Figure~\ref{fig:relaxation_KALJ}(d,e) shows that $O^+(t)$ and $O^-(t)$ not only fluctuate in this regime but are also anticorrelated. Moreover, the fluctuations of $O^\pm(t)$ are correlated with those of $v(t)$ shown in Fig.~\ref{fig:relaxation_KALJ}a. For a closer inspection, a typical enlarged example of the process causing such fluctuations is analyzed in Fig.~\ref{fig:inflection-point}. 
During the decay of $v(t)$, $\hat{\mathbf{v}}(t)$ couples to a stable mode (large $O^+$) with a positive eigenvalue, whereas during the increase, it couples to an unstable mode (large $O^-$) with a negative eigenvalue. The mode remains the same throughout the process, resembling a typical low-energy quasi-localized excitation~\cite{lerner2021low} (see Fig.~\ref{fig:inflection-point}e), but its eigenvalue switches from positive to negative.

The above procedure indicates an event that crosses an inflection point on the energy landscape. At an inflection point, the first derivative of the energy $V$ (i.e., the force magnitude) with respect to the mode coordinate $\xi^*$ attains a minimum, while the second derivative (eigenvalue $\lambda^*$ of the mode) vanishes (see Fig.~\ref{fig:inflection-point}f). When traveling through an inflection point, the system's relaxation first slows down and then speeds up again, much like a car driving over a speed bump on a downhill road. Overall, inflection points introduce bumpy dynamics, reflected in the fluctuations of $v(t)$, superimposed on the overall power‑law behavior of $v(t)$ determined by the caging potential discussed above. However, inflection points are not the essential cause of the slow, power‑law relaxation. Even right at an inflection point, where the velocity reaches its minimum, this minimum does not deviate significantly from the average velocity; inflection points merely add fluctuations to the otherwise smooth power-law decay.

By definition, a saddle point has a vanishing first derivative of the energy (i.e., zero velocity or force). Our single-sample data show no instances where the velocity approaches zero in the power-law regime, suggesting that the GD trajectory remains far from any saddle points. Two further pieces of evidence support the absence of saddle-point influence. First, starting from the neighborhood of an inflection point — where the velocity is minimal — we fail to find any saddle point using the recently developed saddle-dynamics search algorithm~\cite{wu2026hierarchical}. Second, for the 2D MCM, all saddle points are marked in the energy landscape (Fig.~\ref{fig:cage_potential_KALJ}h). Clearly, although the GD trajectories reproduce the power-law behavior, most of them remain far away from any of these saddle points.

We next analyze the statistics of inflection-point events. Figure~\ref{fig:mult_samples}a shows $v(t)$ curves for multiple samples, each relaxed from an independent equilibrium configuration at $T=2$. The fluctuations due to inflection points (IP) become visible only after $\tau_{\rm IP} \approx 0.1$. More quantitatively, we count the cumulative number $N_{\rm IP}$ of inflection-point  events defined as crossings of the eigenvalue $\lambda^*$ (of the eigenmode with the largest overlap with $\hat{\mathbf{v}}(t)$) from positive to negative values through zero (the crossovers with small overlaps $O^{-}(t)<0.25$ are neglected). 
Figure~\ref{fig:mult_samples}b displays the sample-averaged $N_{\rm IP}$, which remains zero below $\tau_{\rm IP}$ and increases beyond this time scale. Additionally, we compute the fraction $f_{\rm LO}^\pm(t)$ of samples where $\hat{\mathbf{v}}(t)$ exhibits a large overlap (LO) with a stable ($+$) or unstable ($-$) instantaneous normal mode at time $t$, i.e., $O^{\pm}(t) > O_{\rm threshold}$.
Figure~\ref{fig:mult_samples}b-inset shows how the onset of inflection-point events depend on $T$.
These analyses indicate that prior to $\tau_{\rm IP}$, the descent proceeds so rapidly in the caging potential that only a few inflection points are encountered. Beyond $\tau_{\rm IP}$, inflection-point events emerge clearly, with the trajectory ``bouncing'' between them---a regime in which the velocity field shows a strong overlap with certain normal modes of the instantaneous Hessian.

\begin{figure}[!htbp]
  \centering
  \includegraphics[width=\linewidth] {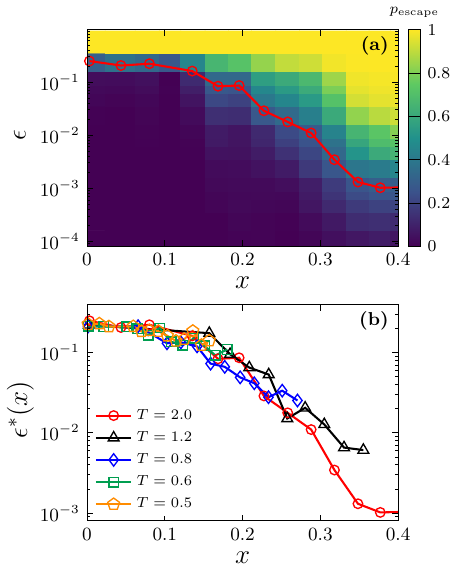}
\caption{{\bf Probability of escaping the basin under perturbations.} 
(a)  The color map indicates {$p_{\rm escape}(\epsilon, x)$} for $T={2.0}$, with the state-following boundary $\epsilon^*(x)$ indicated by the red line.
(b) Collapse of the boundaries $\epsilon^*(x; T)$ for different $T$. 
}
\label{fig:BOA}
\end{figure}

{\bf State following under perturbations.}
The state-following procedure, initially developed within the Franz-Parisi construction~\cite{franz1995recipes}, describes the dynamics in which the system remains in the same metastable glass basin when subjected to perturbations~\cite{rainone2015following, zdeborova2010generalization}. As discussed in the Introduction, mean-field spin glass theory predicts that state-following dynamics occurs only below a threshold temperature $T_{\rm SF}$.
We show below that the situation is essentially different in structural glasses: state following occurs universally at all studied temperatures $T$ as long as the system is near the inherent structure and the applied perturbation is small. There is no characteristic temperature separating the state-following regime from other regimes.

To examine how robust the trajectory is against a perturbation, the instantaneous configuration at time $t$ is perturbed by adding a random Gaussian vector $\epsilon \mathbf{g}$ of magnitude $\epsilon$: $\mathbf{r}_i(t) \rightarrow \mathbf{r}_i(t) + \epsilon \mathbf{g}$, 
where $\mathbf{g}$ is a unit Gaussian vector. 
From each perturbed configuration, GD simulations are performed (without further perturbations) and the resulting inherent structure is compared with the unperturbed one. For each pair of $\epsilon$ and $t$, we simulate {$30$} samples (independent initial equilibrium configurations) and {$50$} realizations for each sample. 
The probability
$p_{\rm escape}(\epsilon, t)$ denotes the fraction of realizations that escape the original basin (see Supplementary Section S4 for the criteria of escaped realizations). {We then recast $p_{\rm escape}$ as a function of $\epsilon$ and $x$, using the relation $x=x(t)$, where $x$ quantifies the distance to the inherent structure as defined in the caging potential Eq.~(\ref{eq:Vcage}). 
The state-following boundary $\epsilon^*(x)$ is defined  as $p_{\rm escape}(\epsilon^*, x)=p_{\rm threshold} = 0.3$ (the red line in Fig.~\ref{fig:BOA}a for $T=2.0$), below which is the state-following regime where the system rarely escapes. As shown in Fig.~\ref{fig:BOA}b, $\epsilon^*(x; T)$ collapse for any $T$ in the studied range ${0.5} \leq T \leq {2.0}$, which covers $T_{\rm onset} \approx 0.7$ and $T_{\rm def} \approx 0.6$.}  


Several important implications follow from the above results. First, state following occurs at any  $T$, provided that both the perturbation and the distance to the inherent structure are small enough to lie below the collapsed boundary in Fig.~\ref{fig:BOA}b. Second, because the boundaries for different temperatures collapse onto a single curve, the GD dynamics are qualitatively similar for all $T$; no characteristic temperature for distinct dynamical regimes can be identified. Third, the long-time (small-$x$) dynamics is robust against perturbations, confirming that the minima are stable (not marginally stable) and that saddle points are not involved.

{\bf Discussions.}
This study establishes a complete physical picture of slow relaxation dynamics in quenched structural glasses. The overall power-law behavior originates from a stiffening caging effect, while fluctuations around this average behavior arise from inflection points on the energy landscape. The glass basins are stable, not marginally stable. Near the bottom of a basin, the system undergoes exponential relaxation governed by the lowest-energy disordered vibrational mode. In this near-bottom region, the dynamics is robust against perturbations and remains within the same basin, thereby resembling  state following.

Our picture differs markedly from all existing ones, especially the mean-field spin glass scenario. In structural glasses, the power-law behavior is independent of saddle points and marginal stability, and no $T_{\rm SF}$ can be defined. The results for the amorphous silica  model support the genericity of the stiffening caging mechanism (see Supplementary Section S5). 
In addition, Supplementary Section S6 shows that a similar exponent $\beta \approx 0.6$ is compatible with the simulation data in the high-$T$ regime for all considered models (KALJ, amorphous silica, soft spheres, and MK). 
These results suggest that our picture in Fig.~\ref{fig:schematic}b applies to general structural glass models.
A likely exception is granular systems near the jamming transition, where marginal stability and hierarchical energy landscapes are well established~\cite{liu2010jamming, charbonneau2014fractal, muller2015marginal, berthier2016growing, jin2017exploring, jin2018stability, wu2026hierarchical}. We therefore expect the GD dynamics to differ substantially in this regime, which would be an interesting direction for future work.

As a final remark, we emphasize that in disordered systems, at least two distinct mechanisms can produce slow relaxation of GD dynamics.
One is many-body confinement, exemplified by the caging effect found here: near confinement walls, acceleration is suppressed because small displacements incur large energy costs. The other is saddle-point slow-down, where the velocity vanishes as the first derivative of the energy goes to zero. 
Although these two mechanisms may coincidentally give rise to similar $v(t)$ behavior with comparable power-law exponents, their underlying differences remain to be elucidated.
Such insight would be valuable for optimizing GD efficiency in diverse applications, including training of artificial neural networks~\cite{huang2025liquid}.

{\bf Acknowledgments.}
We thank Gang Huang and Shuonan Wu for useful discussions. 
The authors acknowledge funding from National Key R\&D Program of China (Grant No. 2025YFF0512000), the China Manned Space Program (Grant No. CMSS-2025-5-P-002),
Wenzhou Institute (No. WIUCASICTP2022) and the National Natural Science Foundation of China (No. 12447101). In this work access was granted to the High-Performance Computing Cluster of Institute of Theoretical Physics - the Chinese Academy of Sciences.

\bibliographystyle{apsrev4-2}
\bibliography{silica}

\clearpage

\centerline{\bf \large Methods}

\vspace{0.5cm}

{\it Appendix A: Simulation model and method for the KALJ model.}
The system consists of an 80:20 binary mixture of
large ($A$) and small ($B$) particles,
under periodic boundary conditions in 3D.
The pair interaction between particles of species $\alpha,\beta\in\{A,B\}$
is given by a truncated and shifted LJ potential:
\begin{equation}
u_{\alpha\beta}(r)
=
4\epsilon_{\alpha\beta}
\left[
\left(\dfrac{\sigma_{\alpha\beta}}{r}\right)^{12}
-
\left(\dfrac{\sigma_{\alpha\beta}}{r}\right)^6
\right]
-
u_{\alpha\beta}^{\rm LJ}(r_{\alpha\beta}^{\rm c}),
\end{equation}
for $r< r_{\alpha\beta}^{\rm c}$; otherwise zero.
The interaction parameters are,
\begin{equation}
\begin{array}{c|ccc}
\alpha\beta & \epsilon_{\alpha\beta} & \sigma_{\alpha\beta} & r_{\alpha\beta}^{\rm c} \\
\hline
AA & 1.0 & 1.0  & 2.5 \\
AB & 1.5 & 0.8  & 2.0 \\
BB & 0.5 & 0.88 & 2.2
\end{array}
\end{equation}
All quantities are reported in reduced LJ
units, with $m_A=m_B=1$, $\epsilon_{AA}=1$, and $\sigma_{AA}=1$.


Equilibrium configurations are prepared as follows. We first create an
face-centered cubic (FCC) lattice at a number density $\rho=1.2$.
The particle types are then randomly assigned to obtain an
 80:20 mixture. The system is  melted using a molecular
dynamics algorithm and then equilibrated at the target
initial temperature $T$.




Starting from each equilibrated configuration, we perform deterministic
overdamped relaxation. The dynamics is implemented using the LAMMPS
\texttt{fix brownian} integrator with the random noise disabled. The  timestep is $
\Delta t_{\rm GD}=10^{-4}$. For all  samples analyzed in this work, the final configuration is converged: the root-mean-square force has decreased below $10^{-11}$ at the end of the overdamped relaxation.\\

{\it Appendix B: Simulation model and method for amorphous silica.}
We study a SiO$_2$ system of 1536 atoms interacting through a hybrid potential that combines long-range Coulomb, short-range Morse~\cite{pedone_new_2006}, and an additional close-range Ziegler--Biersack--Littmark (ZBL) repulsive interaction~\cite{ziegler1985stopping}. The long-range Coulomb interaction is solved using the particle–particle particle–mesh  method with an accuracy of \(10^{-4}\), and the real-space cutoff is set to 8.25~\AA. The simulations are performed using the LAMMPS package in metal units. 

For each pair of atomic species $\alpha$ and $\beta$, with  $\alpha, \beta \in \{\mathrm{Si},\mathrm{O}\}$, the interaction energy is
\begin{equation}
U_{\alpha\beta}(r)=U_{\alpha\beta}^{\mathrm{Coul}}(r)+U_{\alpha\beta}^{\mathrm{Morse}}(r)+U_{\alpha\beta}^{\mathrm{ZBL}}(r).
\end{equation}
The Coulomb term is,
\begin{equation}
U_{\alpha\beta}^{\mathrm{Coul}}(r)=\frac{1}{4\pi\varepsilon_0}\frac{q_\alpha q_\beta}{r},
\end{equation}
where $\varepsilon_0$ is the electrical permittivity of vacuum, $q_\alpha$ and $q_\beta$ are the effective ionic charges. The charges are set to $q_{\mathrm{Si}}=2.4e$ and $q_{\mathrm{O}}=-1.2e$.
The Morse potential is,
\begin{equation}
U_{\alpha\beta}^{\mathrm{Morse}}(r)=
D_{\alpha\beta}
\left[
e^{-2\alpha_{\alpha\beta}(r-r_{0,{\alpha\beta}})}
-2e^{-\alpha_{\alpha\beta}(r-r_{0,{\alpha\beta}})}
\right],
\end{equation}
where $D_{\alpha\beta}$, $\alpha_{\alpha\beta}$, and $r_{0,\alpha\beta}$ are  potential parameters listed in Table \ref{tab:potential_parameters}.  The short-range interaction cutoff is set to 5.5~\AA. 

\begin{table}[htbp]
\caption{Potential parameters used in simulations of amorphous silica \cite{pedone_new_2006}.}
\label{tab:potential_parameters}
\begin{tabular}{cccc}
\hline
$\alpha\beta$& $D_{\alpha\beta}$ 
& $\alpha_{\alpha\beta}$ 
& $r_{0,\alpha\beta}$ 
\\
& (eV) & ($\text{\AA}^{-1}$) & ($\text{\AA}$) \\
\hline
SiO  
& 0.340554 & 2.006700 & 2.100000 
\\

OO   
& 0.042395 & 1.379316 & 3.618701 
\\
\hline
\end{tabular}
\end{table}

The ZBL  potential  accounts for the strong repulsive interaction at extremely short interatomic distances, 
\begin{equation}
U_{\alpha\beta}^{\mathrm{ZBL}}(r)
=
\frac{1}{4\pi\varepsilon_0}\frac{Z_\alpha Z_\beta e^2}{r}\,\phi(r/a)
+ S(r),
\end{equation}
where 
$Z_\alpha$ and $Z_\beta$  are the nuclear charges ($Z_{\rm {Si}}=14$ and $Z_{\rm {O}}=8$), $a$  the screening length, $\phi(x)$  a universal screening function, and $S(r)$  a switching function. 
The screening length is given by
\begin{equation}
a=\frac{0.46850}{Z_\alpha^{0.23}+Z_\beta^{0.23}},
\end{equation}
and the universal screening function is
$\phi(x)=
0.18175e^{-3.19980x}
+0.50986e^{-0.94229x}
+0.28022e^{-0.40290x}
+0.02817e^{-0.20162x}$.
The switching function smoothly ramps the ZBL energy, force, and curvature to zero between the inner (0.8~\AA ) and outer (1.0~\AA ) cutoffs. 


Equilibrated SiO$_2$ configurations at different initial temperatures are generated using the interatomic potential described above. Periodic boundary conditions are employed, and the equations of motion are integrated with a time step of 1.6~fs. The density is fixed at 2.36 g/cm$^3$. Short-range neighbor lists are constructed with a bin size of 2.0~\AA\ and are updated every step with zero delay.
The equilibrated configurations are prepared through a staged thermal protocol in the canonical ensemble  with a thermostat damping parameter of 0.1~ps. A recentering operation 
is applied during the runs in order to remove any overall drift of the simulation box. The target temperature is approached by first melting a crystal silica configuration at a higher reference temperature, then cooling it to the desired temperature, and finally performing an additional isothermal equilibration at the target temperature.




The  GD relaxation dynamics are initiated from the equilibrated parent configurations generated at different temperatures. For each parent state, the system is evolved using the \texttt{fix brownian} integrator in LAMMPS with the stochastic term disabled. 
The equations of motion are integrated with a time step of $10^{-4}$. For each trajectory, the system is evolved for $10^6$ integration steps. After $10^6$ steps, the average force magnitude has typically decreased to the order of $10^{-6}$, indicating that the system has reached a well-relaxed inherent state.\\

{\it Appendix C: Minimum cage model.}
We place $n = 5$ or $6$ neighboring (wall) particles on a circle of radius $R_1 = 1.2$ to form a cage. The angular position of the $i$-th wall particle is given by $\theta_i = 2\pi i/n + \delta\theta_i$, where $\delta\theta_i$ is drawn from a uniform distribution in $[-0.05, 0.05]$. The radial positions are perturbed as $r_i = R_1 + \delta r_i$, with $\delta r_i$ drawn uniformly from $[-0.05, 0.05]$. The mobile central particle interacts with the fixed wall particles through a truncated Lennard-Jones potential,
\[
V_{\rm LJ}(r) = 4\epsilon \left[ \left(\frac{\sigma}{r}\right)^{12} - \left(\frac{\sigma}{r}\right)^6 \right],
\]
with $\epsilon = 1$, $\sigma = 1$, and cutoff distance $r_{\rm cut} = 2.5\sigma$.

For each cage realization, the mobile particle is initialized at a distance $x_0$ from the {origin}, with initial position 
{$\mathbf{r}_0 = \{x_0 \cos\phi,\, x_0 \sin\phi\}$}, where $\phi$ is drawn uniformly from $[0, 2\pi]$. The subsequent relaxation follows overdamped GD dynamics. For each set of parameters, we generate 10 independent cage realizations, and within each realization, the mobile particle is initialized at 10 independent positions.

\end{document}


\title{Supplementary Information\\Relaxation of quenched structural glasses: descent in a stiffening caging potential over inflection‑point `speed bumps'}
\author{Ting Qu}
\affiliation{Institute of Theoretical Physics, Chinese Academy of Sciences, Beijing 100190, China}
\affiliation{School of Physical Sciences, University of Chinese Academy of Sciences, Beijing 100049, China}

\author{Deng Pan}
\affiliation{Institute of Theoretical Physics, Chinese Academy of Sciences, Beijing 100190, China}


\author{Yuliang Jin}
\email{yuliangjin@mail.itp.ac.cn}
\affiliation{Institute of Theoretical Physics, Chinese Academy of Sciences, Beijing 100190, China}
\affiliation{School of Physical Sciences, University of Chinese Academy of Sciences, Beijing 100049, China}
\affiliation{Wenzhou Institute, University of Chinese Academy of Sciences, Wenzhou, Zhejiang 325000, China}

\date{\today}
\maketitle
\onecolumngrid

\setcounter{figure}{0}
\setcounter{equation}{0}
\setcounter{table}{0}
\setcounter{section}{0}
\setcounter{subsection}{0}
\renewcommand{\thesection}{S\arabic{section}}
\renewcommand{\thesubsection}{S\arabic{section}.\arabic{subsection}}
\renewcommand{\thefigure}{S\arabic{figure}}
\renewcommand{\theequation}{S\arabic{equation}}
\renewcommand{\thetable}{S\arabic{table}}
\renewcommand{\figurename}{Fig.}
\renewcommand{\tablename}{Table}

\section{Low-$T$ harmonic-limit regime for the phonon mechanism}
\label{sec:phonon}

\begin{figure}[!htbp]
  \centering
  \includegraphics[width=0.6\linewidth] {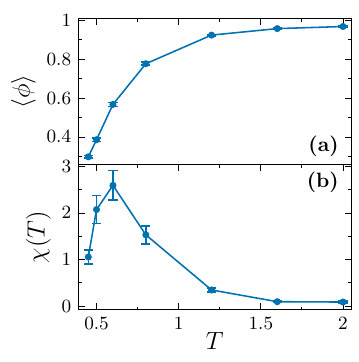}
\caption{ (a) Average concentration $\langle \phi \rangle$ and (b) susceptibility $\chi(T)$ of defects for the KALJ. The peak position of the susceptibility gives {$T_{\rm def} \approx 0.6$}.}
\label{fig:phonon}
\end{figure}

Ref.~\cite{nishikawa2022relaxation} argues that the phonon mechanism applies only at very low temperatures, when localized ``defects'' (particles with large non-affine displacements) are negligible. Quantitatively, this condition corresponds to a vanishing concentration $\langle \phi \rangle$ of defects as $T \to 0$, where $\phi = \frac{1}{N} \sum_i \phi_i$, with $\phi_i = 0$ if particle $i$ does not change its neighbors during GD, and $\phi_i = 1$ otherwise. Correspondingly, the phonon mechanism is expected to hold only well below the susceptibility peak $\chi(T)$, where {$\chi = N(\langle \phi^2 \rangle - \langle \phi \rangle^2 )$}. The peak position of $\chi(T)$ defines $T_{\rm def}$. Fig.~\ref{fig:phonon} shows that {$T_{\rm def} \approx 0.6$} for the KALJ model.

\section{Onset temperature}
\label{sec:onset}

{The onset temperature $T_{\rm onset}$ is estimated from the data of the inherent-structure energy $E_{\infty}(T)$, defined as the averaged  
potential energy  after GD. The low-$T$ ($T<0.65$) and high-$T$ ($T>1.1$) temperature regimes are fitted  separately by linear functions of $T$, and $T_{\rm onset}$ is the intersection of the two lines~\cite{banerjee2017determination}.}
\begin{figure}[!htbp]
  \centering
  \includegraphics[width=0.6\linewidth]
  {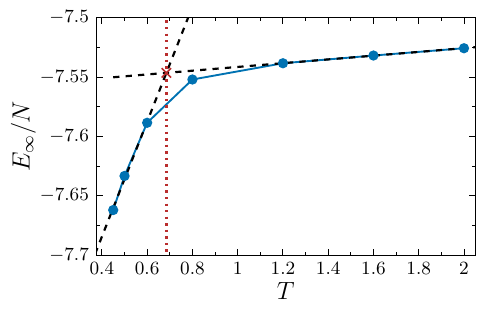}
\caption{{\bf{Onset temperature of the KALJ model.}}
Average inherent structure energy $E_{\infty}$ per particle as a function of $T$. The two lines represent linear fitting to low-$T$ and high-$T$ data, whose intersection gives $T_{\rm onset}\approx0.7$. Data are obtained for $N=256$.
}
\label{fig:onset}
\end{figure}


\section{Logarithmic coarsening }
\label{sec:coarsening}

\begin{figure}[!htbp]
  \centering
  \includegraphics[width=0.7\linewidth] {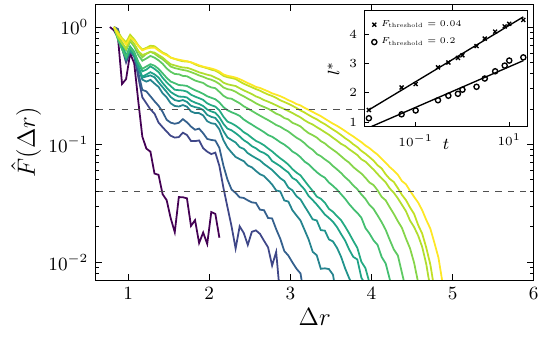}
\caption{ {\bf Logarithmic coarsening in the KALJ model ($N=2048$).}
{Main panel:} $\hat{F}(\Delta r)$ for $10^{-2} \leq t \leq 20$ (curves from left to right correspond to increasing time). 
{Inset:} $l^*$ as a function of $t$ for two threshold values $F_{\rm threshold} = 0.04$ and $0.2$. Solid lines represent fits to $l^* = c_1 + c_2 \ln t$, with $c_1=3.3, c_2=0.41$ for $F_{\rm threshold} = 0.04$ and $c_1=2.2, c_2=0.30$ for $F_{\rm threshold} = 0.2$.}
\label{fig:coarsening}
\end{figure}

In Ref.~\cite{chacko2019slow}, the coarsening length $l^*$ is characterized by the correlation length extracted from the spatial correlation function of thresholded particle speeds. Following this idea, we define a correlation function
\(
F(\Delta r) = \big\langle \bigl[\chi_v(r) - \langle \chi_v \rangle \bigr] \bigl[ \chi_v(r+\Delta r) - \langle \chi_v \rangle \bigr] \big\rangle,
\)
where $\chi_v = 1$ for particles in the top $10\%$ of the speed distribution and $0$ otherwise. The normalized function is $\hat{F}(\Delta r) = F(\Delta r)/F(1)$, where $\Delta r = 1$ (in LJ units) corresponds approximately to the first peak of the pair correlation function $g(\Delta r)$ (see Fig.~\ref{fig:coarsening}). The coarsening length $l^*$ is then defined by $\hat{F}(l^*) = F_{\rm threshold}$. The inset of Fig.~\ref{fig:coarsening} shows that, regardless of the threshold $F_{\rm threshold}$, $l^*$ grows logarithmically slowly: $l^* \sim \ln t$.

\section{Criteria for escaping a basin under perturbation}
\label{sec:perturbation}

For a configuration \({\bf r}(t_a)\) at time \(t_a\), we generate several independent random  Gaussian perturbations with an amplitude \(\epsilon\). We then compare the inherent structure obtained from the perturbed configuration with the reference inherent structure obtained from the unperturbed trajectory. The comparison is made using two independent diagnostics: the relative {root-mean-squared displacement} (RMSD) between the two final configurations and the difference in inherent-structure energy per particle,
\begin{equation}
d_{\rm IS}
=
\left[
\frac{1}{N}
\sum_{i=1}^{N}
\left|
{\bf r}_{{\rm IS},i}
-
{\bf r}^{\rm ref}_{{\rm IS},i}
\right|^{2}
\right]^{1/2},
\end{equation}
\begin{equation}
\Delta e_{\rm IS}
=
\frac{
\left|
E_{\rm IS}
-
E^{\rm ref}_{\rm IS}
\right|
}{N}.
\end{equation}
In addition to the comparison of the final configurations, we also compare the configuration $\{\mathbf r_i(t) \}$ on the perturbed trajectory at time $t$ with that $\{\mathbf r_i^{\rm ref}(t) \}$ on the unperturbed trajectory at the same time, 
\begin{equation}
d(t)
=
\left[
\frac{1}{N}
\sum_{i=1}^{N}
\left|
\mathbf r_i(t)
-
\mathbf r_i^{\rm ref}(t)
\right|^{2}
\right]^{1/2}.
\end{equation}
By definition, $d(0) \sim \epsilon$ and $d(\infty) = d_{\rm IS}$.

The scatter plot of $\Delta e_{\rm IS}$ versus $d_{\rm IS}$ reveals three clearly distinct populations. 
(i) The first population (\emph{convergence}) collapses to the numerical floor, where both the RMSD and the energy difference lie within the range of numerical uncertainties. In this case, the perturbed system converges back to the unperturbed inherent structure.
(ii) The second population (\emph{escape}) comprises realizations that leave the unperturbed basin as a consequence of the GD process. We adopt $d_{\rm IS} > 10^{-2}$ as the criterion for classifying 
$p_{\rm escape}$. Using energy-based criteria would yield comparable classifications.
(iii) A third, intermediate population (\emph{remain}) falls within the range $10^{-6} < d_{\rm IS} < 10^{-2}$. These cases predominantly occur when the perturbation is applied at a very small distance to the reference inherent structure ($x < 0.07$). Notably, multiple inherent structures (or sub-basins) exist at the bottom of each glass basin. When a perturbation is applied near the bottom of a basin, it shifts the system into an adjacent inherent structure. 
As a result, the perturbed system remains within the same glass basin but ends up in an inherent structure different from the unperturbed one. Our GD data (see Fig.~5a of the main text) suggest that all sub-basins are stable, because $v(t)$ always vanishes exponentially at large times.

\begin{figure}[!htbp]
  \centering
  \includegraphics[width=\linewidth] {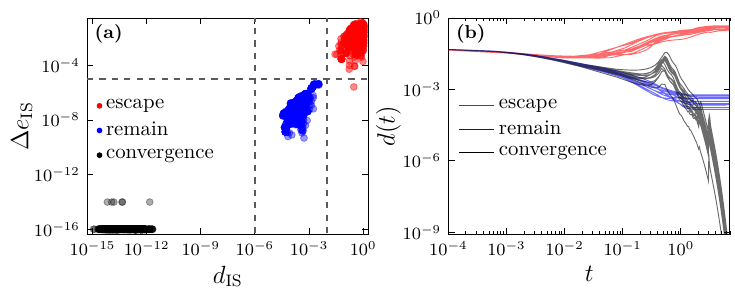}
\caption{
{\bf Three groups of GD dynamical behaviors after perturbation.} 
(a) Scatter plot of $\Delta e_{\rm IS}$ vs. $d_{\rm IS}$ after perturbation. Vertical dashed lines denote our criteria: $10^{-6} < d_{\rm IS} < 10^{-2}$ ($p_{\rm remain}$) and $d_{\rm IS} > 10^{-2}$ ($p_{\rm escape}$). The horizontal line marks the alternative energy-based threshold $\Delta e_{\rm IS} > 10^{-5}$ for escaped cases.
(b) Typical $d(t)$ trajectories for the three groups, obtained with $\epsilon = {0.05}$ at $x = 0.4$ (red, escape), $0.08$ (black, convergence), and $0.0008$ (blue, remain). Data correspond to $T = 2$.
}
\label{fig:perturbation}
\end{figure}

To further illustrate the key dynamical differences among the three populations, we plot representative $d(t)$ curves for each group (see Fig.~\ref{fig:perturbation}b). All perturbed trajectories are initialized with the same perturbation magnitude $\epsilon$, so the $d(t)$ curves start from comparable values at $t = 0$. Their subsequent evolution, however, differs markedly. (i) For the convergent trajectories (black), $d(t)$ rapidly decays to zero at large $t$, indicating that the perturbed system converges back to the same inherent structure as the unperturbed one.
(ii) For the escaped trajectories (red), $d(t)$ increases monotonically with time, showing that the system is driven away from the unperturbed trajectory during GD and eventually settles into a different basin. The GD procedure plays an essential role here, as it amplifies the initial perturbation and is thus the mechanism that pushes the system into a distinct basin.
(iii) For the remaining trajectories (blue), $d(t)$ initially decreases and then levels off to a nearly constant nonzero value at large times. The initial decrease reflects the fact that both systems lie within the same basin, while the plateau at long times indicates that they ultimately terminate in two different sub-basins within that same glass basin.

\section{Gradient descent dynamics in amorphous silica}
\label{sec:SiO2}

\begin{figure}[!htbp]
  \centering
  \includegraphics[width=\linewidth] {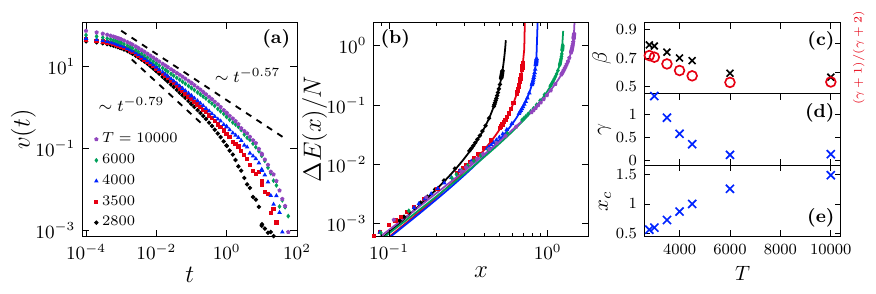}
\caption{ {\bf  Gradient descent dynamics and the effective caging potential in simulated amorphous silica.}
{(a) Average velocity $v(t)$ for different temperatures $T$, with power-law fits yielding $\beta$ (see panel c). 
(b) Average energy $\Delta E(x)$ per particle at different $T$, fitted to Eq.~(4) of the main text (solid lines), with the  fitting parameters $\gamma$ and $x_c$ given in (d, e). 
(c) Comparison between $\beta$ from the fits in (a) (black crosses) and $(\gamma+1)/(\gamma+2)$ (red circles), where $\gamma$ is obtained from the fits in (b). } 
}
\label{fig:relaxation_silica}
\end{figure}

{We perform simulations for amorphous silica glasses ($\rm {SiO_2}$) of $N=1536$  particles to examine the mechanism discussed in the main text. The simulations are carried out at different initial temperatures ranging from $2800\,\rm K$ to $10000\,\rm K$. 
As shown in Fig.~\ref{fig:relaxation_silica}, the GD dynamics can also be explained by an effective caging potential, suggesting that this mechanism  is insensitive to the microscopic details of the glass model.
}

\section{Comparison of $v(t)$ for different structural glass models at the highest reported temperature}
\label{sec:model_comparison}
\begin{figure}[!htbp]
  \centering
  \includegraphics[width=0.6\linewidth] {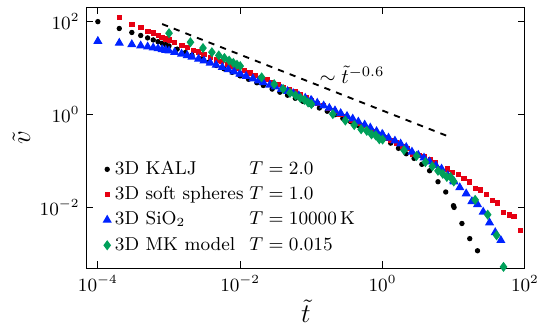}
\caption{ {\bf Comparison of $\tilde{v}(\tilde{t})$ for different models.} For each model, the data are collected at the highest reported $T$, which is in the temperature range where the exponents are independent of $T$. Data for $3\rm D$ soft spheres and the 3D MK model are obtained from Ref.~\cite{nishikawa2022relaxation}. Data for the 3D KALJ and the amorphous silica are obtained in this study.
}
\label{fig:compare_models}
\end{figure}

To compare the high-temperature decay behaviors across different models, we rescale $v$ and $t$ by model-dependent factors ($v_0$ and $t_0$), as $\tilde{v} = v/v_0$ and $\tilde{t} = t/t_0$, to have best collapse of the data in Fig.~\ref{fig:compare_models}. This alignment shows that the exponent $\beta$ in the high-$T$ regime depends weakly on the model.


\bibliographystyle{apsrev4-2}
\bibliography{silica}